# URBAN ECONOMICS OF MIGRATION IN THE CITIES OF THE NORTHEAST REGION OF BRAZIL

**Denise Cristina Bomtempo**
Universidade Estadual do Ceará-UECE
denise.bomtempo@uece.br

**Abstract**

The focus of this text is to discuss how geographical science can contribute to an understanding of international migration in the 21st century. To this end, in the introduction we present the central ideas, as well as the internal structure of the text. Then, we present the theoretical and methodological approach that guides the research and, in section three, we show the results through texts and cartograms based on secondary data and empirical information. Finally, in the final remarks, we summarize the text with a view to contributing to the proposed debate.

## Introduction

In Brazil, new international migration routes have been mapped and materialized in cities of different sizes and functions in all regions of the country. In the Northeast of Brazil, since the 1990s, the development of regional policies has attracted national and international investment in all sectors of the economy. Among the consequences of these investments is the intensification of migratory movements involving different profiles of subjects.

Among the cities in the Northeast with the highest volume of international migration are Salvador, Recife and Fortaleza. In this text we intend to present the migration of Africans settled in the cities of the Northeast region of Brazil, with a focus on the city of Fortaleza/CE, who contribute to the emergence of an urban economy of migration.

## Theoretical and methodological contributions to a geographical reading of migration in the 21st century

Our approach in explaining migration is not centered in the individuals themselves, but in space and territories which have dynamics shaped by structural and conjunctural forces that interfere in the trajectory of those involved in migration. It is worth pointing out that these dynamics are not materialized homogeneously, since there are multiple and coexistent migratory flows.

Unlike a purely statistical analysis that places the subject in separated spaces and times (departure and arrival). According to Sayad (1998), migration is read by taking into account that the migrant is the subject with the capacity to articulate space, time and people through their path, trajectory and circularity. This geographical perspective is not based solely on an orthodoxy of numbers, but on the elements which constitute the territories and the spatial and social relations of the subjects investigated.

In order to deal with the complexity of migration in the 21st century, we defined operational methodological procedures that include the choice of variables and indicators found in statistical databases, but also we make use of unique empirical qualitative research sources linked to the empirical scale of reading the object.

The statistical bases used in the first stage of the investigation were not sufficient to build an apparatus which could read the object. For this reason, we adopted qualitative procedures to allow closer involvement with the investigated group. When it comes to international migration, we organized the information regarding the subjects' nationality. Using the nationality variable as a starting point, we tried to verify: a) territorial origin, year of decision to migrate, migration route (means of transport, support from social, family and institutional networks); b) reasons that led to migration, educational background, gender, age, single or composite migration (family group); c) housing, social and educational integration: housing conditions, access to social and healthcare programs, school (in the cases of families with children); d) economic activities carried out before, during the migration route and at the time of the interview.

Regarding the economic activities, the aim is to analyze if migrants are involved in: a) the formal market; b) informal activities; c) whether they have the capacity to invest and structure their own businesses.

In this text we focus our analysis on the economic activities carried out by international migrants from African countries. To this purpose, we used elements of the methodology aforementioned and we have as one of our references the work developed by Milton Santos, systematized in *O espaço dividido* (1979) and the researches developed by Silveira (2015) and Montenegro (2012) which contribute to read the urban economy of countries that gone through processes of colonial exploitation associated with slavery and still have an unequal capitalist economic structure based on the coexistence of formal and informal activities (that increase in periods of structural and conjunctural crises).

In Santos' words (1979), "[...] the existence of a mass of people with very low salaries or living from occasional activities, alongside a minority with very high incomes, creates a division in urban society between those who can have permanent access to the goods and services on offer and those who, having the same needs, are unable to satisfy them. This creates both quantitative and qualitative differences in consumption. These differences create circuits of production, distribution and consumption of goods and services" (p. 37).

The urban economy of "underdeveloped" countries, according to Santos (1979), forms two circuits of the urban economy (upper and lower). These two circuits coexist, they are complementary and articulated. They manifest themselves in the city and form two subsystems (of the urban system) "[...] the upper or modern circuit and the lower circuit". "[...] the upper circuit is originated from technological modernization and its most representative elements today are the monopolies. The bulk of their relations take place outside the city and region that house them, and their setting is the country or abroad" (p. 22). "[...] the lower circuit is made up of small-scale activities and is especially concerned with the poor population; on the other hand, it is well rooted and maintains privileged relations with its region" (p. 22). Yet, [...] "the lower circuit comprises traditional manufacturing activities, such as handicrafts, as well as traditional transportation and services" (p. 24). Still according to Santos (1979), it is in the economies of "underdeveloped" countries that it is possible to notice the existence of the "marginal upper circuit" "[...] permeated by activities with a predominance of characteristics of the upper circuit, but also have characteristics of the lower circuit" (p. 42, 43).

By analyzing the two circuits, we can understand the city as a whole and, therefore, the process of urbanization - in its generalities, particularities and singularities. In its relationship with the world and in its internal relations with the region.

Bomtempo's work (2019, 2020) showed that international migrants in Brazilian cities carry out activities that are linked to the upper, upper marginal and lower circuits of the urban economy. Those who arrive through forced migration conduct activities to ensure their survival, which are especially linked to the lower circuit of the urban economy. However, we now realize migratory movements coexist and with this, there are groups of migrants entering the national territory who are part of a group of migrants with an investment profile and who have resources acquired in the territory of origin (Bomtempo, 2020).

Through the international migrants present in Brazilian cities, we can notice a greater diversity of activities that need to be mapped, analyzed and, if possible, classified from an economic and spatial point of view, in terms of belonging to the upper, upper marginal or lower circuits. Besides, it is interesting to verify if these activities can contribute to the emergence of an urban economy driven by migration - the "urban economy of migration" (Bomtempo, 2020).

In verifying the emergence of an urban economy of migration, it is necessary to check the specific demands of these migrants, both for the development and continuity of activities, and also to reproduce themselves as social subjects in the territory of migration, as pointed out by Bomtempo, 2020; Bomtempo and Sena, 2021; Araújo and Bomtempo (2022).

**Migration across Brazilian territory in the 21st century: a look at the Northeast region**

From 1990 to the current days, on a spatial and temporal scale, migratory flows have coexisted in Brazil. What is new regarding international migration is the existence of a profile of migrants from border countries, Latin American countries, African and Asian countries that make up the Global South (Baeninger, 2018).

On a global scale, the tightening of migration policies, economic and political instability and the metamorphosis of the labour market with neoliberal characteristics led to the intensification of blocking, closures, militarization and standardization of borders by national agents. As a result, migrants had to seek alternative migratory routes, one of which is Brazil.

Brazil's political and economic stability (2003-2014), the dispersion of economic activities across the country, social programs to reduce inequalities, investments (internal and external) in several sectors of the economy and the revision of the Migration Law (n. 13.445/2017), gave Brazil a centrality at the directions of existing flows at the beginning of the 21st century.

The factors listed (World and Brazil) allowed an increasing of the volume of migration and refuge seekers on a global scale, in Latin America and in Brazil. In the latter, there has been a diversification of the profile in terms of the reasons for migration (we noticed a massive presence of forced migrants, investor migrants, migrants for training purposes - studies). It is possible to see that new migratory routes have been mapped in Brazilian territory, especially involving the border and cities of different sizes and functions in the national urban network, and specifically in the North and Northeast of Brazil.

With regard to the presence of international migrants in the Northeast region, it is interesting to highlight, according to Pereira Júnior (2012), that since the 1990s the development of regional policies which aimed reducing imbalances has attracted national and international investments and investors in sectors linked to tourism, commerce, industry, services and agro-industry. Ceará is one of the northeastern states that most ensured the dynamization of economic activities through regional development policies. One of the consequences of this economic dynamization was the intensification of migratory movements on different scales and with different profiles. Among the cities in the Northeast with the highest volume of international migration are Salvador, Recife and Fortaleza. The latter experienced international migration more recently, according to data from SINCRE (2021).

In summary, in the first decades of the 21st century, the presence of companies and migrants with an investment profile in the cities of the country's Northeast region was guaranteed by, among other factors, the revision of the Brazilian Migration Law from 2003 to 2016 and approved with revisions that were not part of the discussions and agreements previously signed in 2017 (minimum investment 2009 - 2015; minimum investment 2015 to the present day R$ 500,000.00); f) creation of the National Statute for Micro and Small Enterprises (BOMTEMPO, 2020).

The presence of a migrant population with an investment profile from the countries of the global North and South has ensured the emergence of an economy centered on migrant capital and labor. Bomtempo (2020) named it "urban economy of migration", whose characteristics include: a) investment capital coming from the migrant themselves or acquired from national financing agencies in the territory of migration; b) employment of family labor, of people from the country of origin and workers linked to the territory of migration; c) marketing of products that refer directly or indirectly to the country of origin; d) acquisition of products for marketing through networks mainly structured by agents who have the same territorial origin as the migrants; e) a production or marketing circuit configured by a continuous, contiguous and trans-scalar network; f) the presence of innovations that can be and/or from the point of view of production, organization, marketing and relations between those involved in economic activities; g) usage of online social media to promote and sell products; h) participation in groups, institutions and social media involving people from the same territory of origin with the aim of debating issues linked to the territory of migration and also often cooperating on supplies, ideas and different logistics; i) activities that are part of the "upper, upper marginal and lower circuits of the urban economy".

In the Northeast region of Brazil, the state of Ceará has the highest number of international migrants (forced migration, migration for study and investment). Investments from migrants' capital are mainly carried out in Fortaleza, in municipalities on the coast and in the interior of the state of Ceará, which play a central role in Ceará's urban network. Cartogram 1 shows that the spatial distribution of international micro-investors with permanent visas (2019) according to SEBRAE is not homogeneous and that the Southeast of the country concentrates the largest volume of investments, followed by the South, Northeast, Midwest and North.

Regarding the Northeast region (Cartogram 2), it is important to consider that this region has only recently been evidenced in terms of the existence of direct investments (domestic and international), which is why the investigation into this empirical section is justified. Among the states in this region, Bahia, Pernambuco and Ceará are the most prominent in terms of the volume of investors. Ceará stands out among the three in terms of the diversity and origin of its investors.

**Cartogram 1**                   **Cartogram 2**

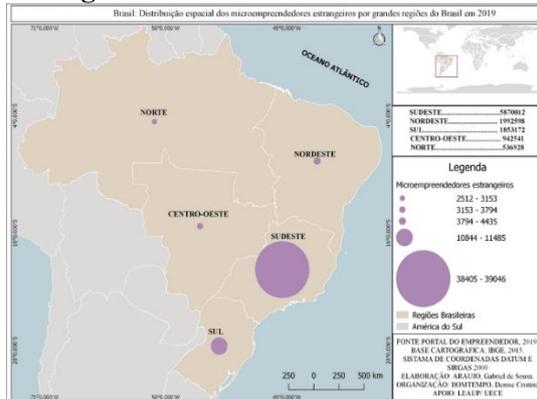 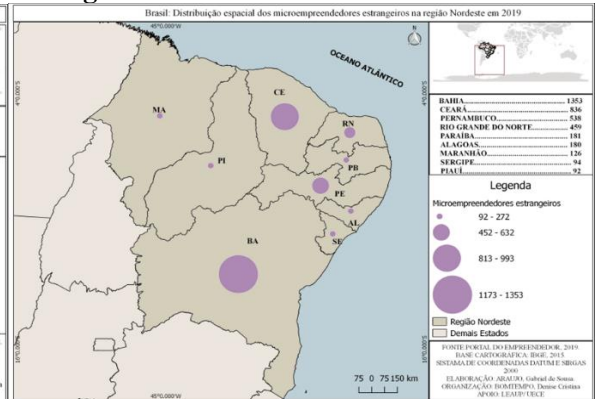

**Cartogram 1 e 2**: Territorial distribution of micro investors in the Northeast region of Brazil (2019).

Concerning the origin of the international migrants who invest in the Northeast region of Brazil. On the one hand, Europe takes place as the continent of origin for the largest number of micro-investors. Italy, Portugal, France, Spain and Germany stand out as the countries of origin of these migrant investors. On the other hand, the countries of Oceania (Australia and New Zealand) have the lowest number of migrant investors in the Northeast region of Brazil (between 10 and 20).

**Cartogram 3**                   **Cartogram 4**

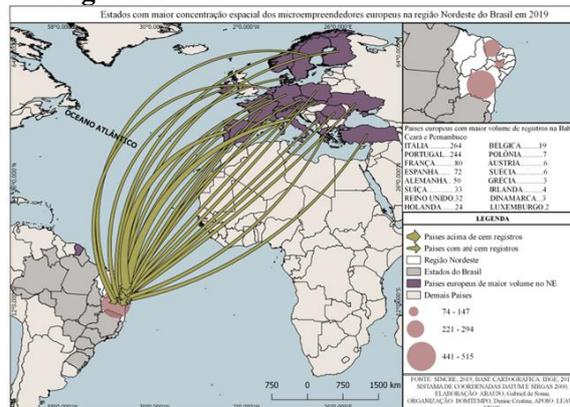 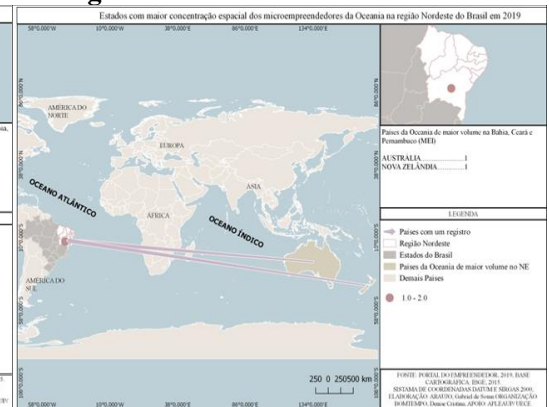

**Cartogram 3 e 4**: Territorial origin of micro investors from Europe and Oceania in the Northeast region of Brazil (2019).

The countries that compose the block of Latin American countries also play an important role in terms of the origin of migrant investors in the northeast of Brazil, with special emphasis on Argentinians, Colombians, Venezuelans, Chileans and Cubans, as shown in Cartogram 5.

Considering the origin of the African investors, we can see that a significant proportion come from regions that are part of the Maghreb, as well as the countries in the west and south of the continent. It is interesting to notice that a considerable number of these countries, like Brazil, were Portuguese colonies and therefore speak the same language, as is the case with Guinea-Bissauans, Cape Verdeans and Angolans. Among the states with the highest number of investors are Bahia/BA, Pernambuco/PE and Ceará/CE.

| **Cartogram 5** | **Cartogram 6** |
|---|---|
| 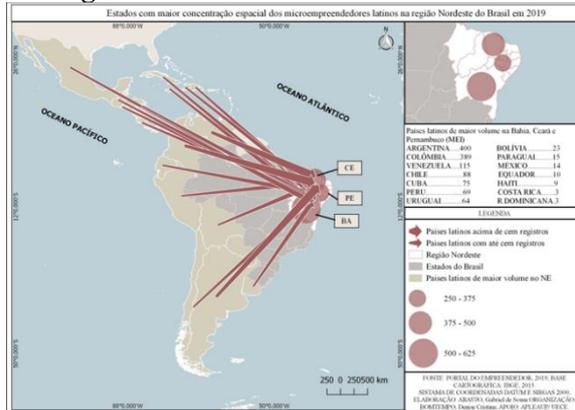 | 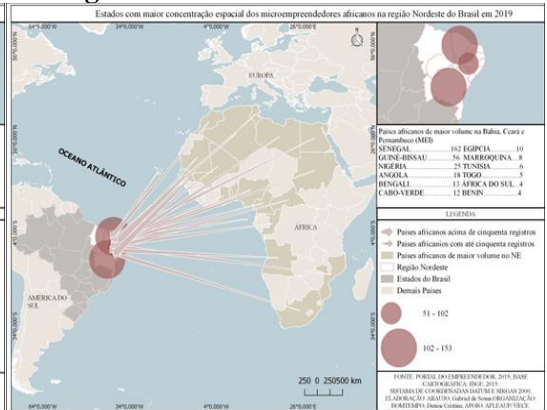 |

**Cartogram 5 e 6**: Territorial origin of micro investors from Europe and Oceania in the Northeast region of Brazil (2019).

In respect to Asian investors, China, Japan, India, North Korea and Lebanon are the countries that are notable in terms of the origin of migrant investors, as can be seen in Cartogram 7. Bahia/BA, Ceará/CE and Pernambuco/PE are the states with the highest concentration of these migrant investors.

**Cartogram 7**

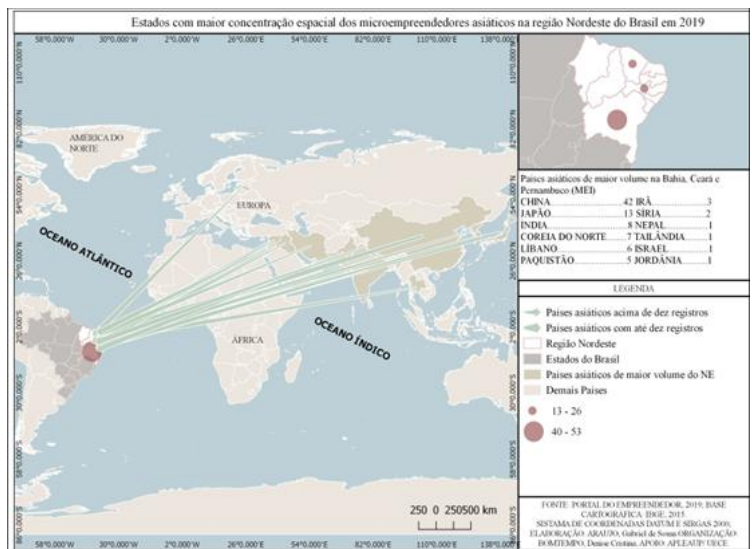

**Cartograma 7**: Territorial origin of Asian micro investors in the Northeast region of Brazil (2019).

International migrants with an investor profile are present in the cities of the Northeast region of the country, carrying out their work activities, not without conflicts, in order to guarantee their existence, their permanence and the possibility of continuing to exist. According to Bomtempo (2020), in the current period, the directioning of migratory flows, especially the international ones, which materialize and intertwine in Brazilian territory is different from other periods from the perspective of the causes, direction, spatial scale, temporality of migration and the profile of the people on the move.

In relation to the causes that lead to the existence of flows that materialize in Brazil, the economic variable is one of those that allows us to follow a path to interpret the migratory phenomenon, given that the configuration and content of migrations are strongly articulated to the dispersion of economic activities linked to industry, commerce, services and agribusiness in the national territory. Meanwhile, the Northeast region of Brazil, especially the state of Ceará, which for decades was source of migrants, is currently a welcoming territory for internal and international migration.

As far as international migration is concerned, on arriving in the territories of migration, these migrants get involved in various work activities, mainly through the intermediation of networks set up during migration. As a result, they produce the formal and informal city, boost the urban economy and foster new territorialities on the scale of the city and the region where they live.

In the state of Ceará, the investments made by migrants of other nationalities are respectively: a) large investments: industry; agribusiness; land purchase; development; construction industry; real estate; hotels; imports and exports. b) small and medium investments: urban economy - commerce and services. Investments from migrants' capital are mainly made in Fortaleza, in municipalities on the coast and in the interior of the state, which play a central role in Ceará's urban network.

The construction of networks is an important factor in migrants' ability to initially keep in the place of migration and weave their relationship with themselves, with the other and with the territory. When these migrants stay, they are not alone, but organized into migratory networks in which they articulate people and territories with the perspective of building their multiple territorialities.

When they move to cities in Ceará's urban network, migrants structure their networks, mostly made up of those who have the same territorial origin, they dedicate themselves to work as skilled and unskilled workers, they make investments, creating an urban economy made up of migrant capital and they create multiple territorialities based on their choice of where to live, work, consumption and leisure. Therefore they configure a differentiated use of city space and territory in the city, as well as the coexistence of different territorialities.

The activities carried out by migrants whose territorial origins are linked to the global South allow us to discuss dynamics that are attached to the investments of micro and small scale made by these migrants in the territory of migration, which give rise to the "urban economy of migration" (Bomtempo, 2020) as a strategy for economic and social reproduction in the territory of migration. As examples, we can mention the activities carried out by migrants from African countries.

# Urban economy of migration in the cities of the Northeast region of Brazil developed by African migrants

According to Bomtempo and Sena (2021), African migrants have been present in Ceará since 2012, when people from various countries on the continent were registered as refugees, skilled and unskilled workers, university students and investors.

Among the investments made by African migrants in Ceará, those involved in the upper, upper marginal and lower circuit of the urban economy stand out. Among the activities, those bounded to construction industry stand out; commerce (restaurants, stores, independent sales: on the streets and at fairs); services (hotels, hostels, gyms, vehicle rental); industry (engineering and information technology); import and export activities. The majority of investments are in the city of Fortaleza, municipalities in its Metropolitan Region and coastal municipalities, as shown in Cartogram 8.

**Cartogram 8**

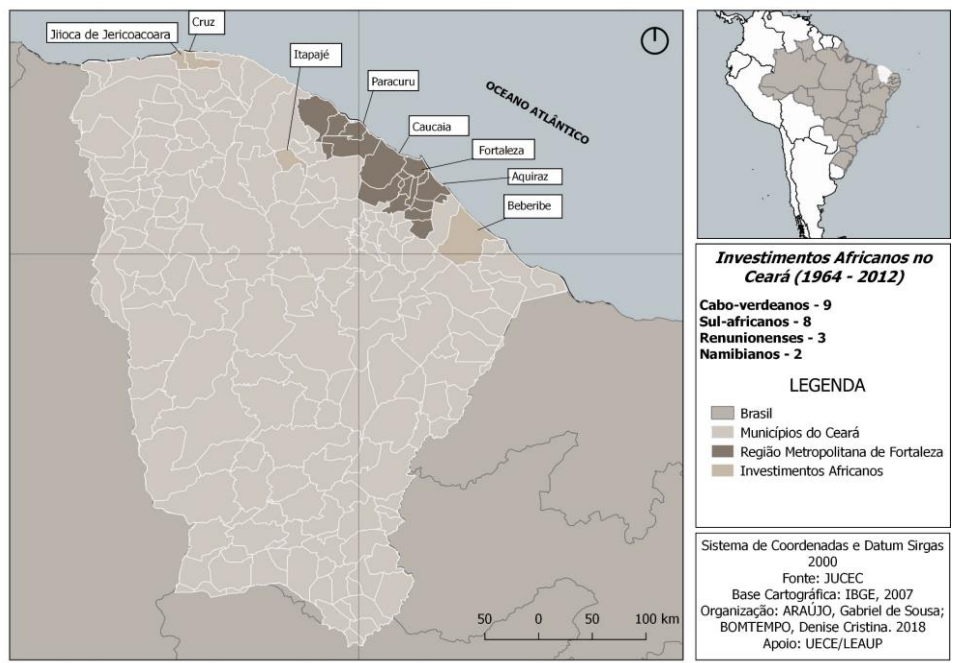

**Cartogram 8**: African investments in the state of Ceará/Brazil (1964 - 2012).

In this text, we present three examples of an urban economy of migration that materializes on the scale of the city of Fortaleza. As an example of an urban economy of migration attached to the lower circuit of the economy, we present the marketing of products implemented by the Senegalese. Next, an example of the urban economy of migration supported by activities linked to the marginal upper circuit is the production of ethnic clothing by a migrant from Guinea-Bissau and finally, as an example of the urban economy of migration convergent to the upper circuit of the urban economy, we take the investments made by migrants representing an internet submarine cable company.

The urban economy of migration supported by activities in the lower circuit of the urban economy is mostly associated to the commerce of low-cost products sold by street vendors in the city center of Fortaleza, on the beaches with a high concentration of tourists (Praia do Futuro and Beira Mar) and in the street markets of the municipalities in the Metropolitan Region of Fortaleza. These migrants sell decorative accessories (jewelry and sunglasses), coconut water, clothes and beachwear accessories (cangas painted with elements of African culture). They are mostly Senegalese, young men and women.

This profile of migrant lives in peripheral neighborhoods in the city of Fortaleza and municipalities in the Metropolitan Region. Most of them share their homes with people of the same territorial origin and use their living space to store the products they sell.

In Brazil, they migrate around Brazilian states that are not part of the Northeast region of Brazil and constantly return to their country of origin (both those who have regularized documentation according to Brazilian law and those who do not - they constantly re-enter Brazil so as not to be irregular (from a documentary point of view) and thus are at risk, as can be seen in the testimony of one of the interviewees:

> "[...] I'm Senegalese and I arrived in Brazil on December 24, 2016. I came to earn money. I used to watch the soap operas and I liked here. My two brothers live in Spain. Brazil isn't what I expected... it's beautiful by nature, but I thought it would be more beautiful. I didn't find it difficult to start working here. I always wanted to work alone. In Senegal I worked driving trucks, selling fish in other cities. Here in Brazil I have my documents organized (SSN, residence permition, everything). I came to Brazil alone and now I'm going back to my country to stay there for four months. I live in Cumbuco (Icaraí) and I come here to work (São Bento street market - Cascavel/Metropolitan Region of Fortaleza, which takes place weekly on Sundays. Cascavel is 83 km from where I live). I work from Sunday to Sunday. I sell in the center of Fortaleza, at street markets and at festivals in the countryside. The products I sell I get from a friend who is also Senegalese and lives in São Paulo. I travel everywhere. In January I go to the beaches in the south of Brazil. When it gets slow, around April, I come back here. In July I go to Senegal and stay for four months. I think it's good here. I've been here for three years and I have a car. I've done a lot in Senegal with my work here. If I'd been there, I wouldn't have been able to, you know? I don't go out much, you know? I work to a live, you know?" (Dio - fictitious name. Interview given at the Cascavel street market, July 2021).

Since the conversation with Dio (July 2021), it has been possible to see in the landscape of the street markets in the Metropolitan Region of Fortaleza, in the street vending in the city downtown and in the beach environments (Beira Mar and Praia do Futuro) of Fortaleza, that the number of street vendors, both Brazilian and of other nationalities, has increased. From the interviewees' testimonies, it can be seen that informal activity has become an alternative to the economic, political, social and sanitary instabilities associated with the extreme right-wing profile of the national government (2018 - 2022) and the problems arising from the lack of federal management of the Covid-19 pandemic.

Among Senegalese migrants, the growth of activities attached to the lower circuit of the urban economy is sustained by "made in China" goods purchased by a network of suppliers (also Senegalese) distributed to sellers of the same territorial origin. Contacts for distribution and sale are established through informational social media and the goods are transported by land and air companies.

This technical and social network allows the development of trade fostered by the Senegalese in the cities and contributes to the emergence and permanence of an urban economy of migration associated to the lower circuit of the economy, since these are activities executed by migrants of the same territorial origin who store, distribute and sell products articulated by multiple scales in cities of different sizes and functions in the Brazilian urban network.

Moreover, the production and marketing of clothing by African migrants allows us to identify elements of an urban economy of migration paired to the upper marginal circuit of the urban economy, as we can see in Joana's (2021) testimony. Our interviewee, Joana, who was born in Guinea-Bissau, arrived in Brazil in 2008 to study after being selected to study. Brazilian public universities, through an agreement between the Brazilian federal government (Luiz Inácio Lula da Silva), and universities in Portuguese-speaking countries in Africa, made a series of opportunities of exchanges, agreements, exchanges of experiences and professional training for students of African origin were established (Sena and Bomtempo, 2018).

When she arrived in Fortaleza, Joana lived in several neighborhoods, all of them far from her place of study, so she made multiple journeys by public transport in the city. During the time she studied at the public university, she wore ornaments and clothes from her country of origin. According to Joana, *"some people thought my clothes were different and a bit patterned, but a lot of people liked them and started asking me to make them. At first I sold the sewn clothes I brought, but then I started sewing"* (Joana, December 2021).

Regarding the statement, it is possible asserting that the clothes produced and sold by Joana were an element of articulation and a symbol of integration between the migrant population and the local population. Because she lives in a neighborhood on the periphery of Fortaleza (Antônio Bezerra), and because orders for clothes made with African-print fabrics have increased, Joana has not found it difficult to hire seamstresses, since, according to Nobre (2018), Fortaleza is central to the clothing industry and industrial production is executed in a fragmented way, mainly by seamstresses who work at their homes and earn for each piece sewn.

Since she arrived in Fortaleza to study in 2008, Joana started selling some African clothes she had brought back in her suitcase. In 2013, she registered as a micro-investor, created a company with a brand, did a graphic design project and started employing a seamstress who works with a formal contract. Since 2018, she has hired professional service providers, such as models, saleswomen, an accountant and a photographer. Whenever possible, she prefers to hire women (African and Brazilian).

Joana does multiple activities. She designs clothes and jewelry, helps with the sewing, sells in her physical store, which is also located in a neighbourhood on the periphery of Fortaleza (close to where she lives), organizes sales on social media, sends the clothes she sells by carrier and participates in collectives, such as the Network of African Women Investors.

As far as sales are concerned, Joana said that after she started promoting her products on social media and using models to display the garments, the orders for African clothing increased enormously, both from Africans and Brazilians living in many cities across the country. Among her customers are those living in Bahia, São Paulo and Brasília. However, she also reported that she sells to cities in the states of Rio Grande do Sul and Minas Gerais. Sales are made in cash and by credit and debit card.

To acquire the main raw material - the fabric (called ankara, capulana or pano), Joana revealed she only gets the goods through members of her family network. She has a Senegalese friend who lives in Guinea-Bissau and buys the fabric for her. "*When I need fabric, I call my friend and we arrange a day and time for him to go to the market. Then we have a video call, he shows me the prints, I choose and we decide on the quantity together. He sends the fabrics by ship to my cousin who lives in São Paulo. He picks up the fabrics and sends them by bus to me here in Fortaleza*" (Joana, December 2021).

From Joana's statement, it can be seen that the activity she carries out involves people linked to her territorial origin, as well as people with whom she has come to have relationships (through the economic activity she carries out) in the territory of migration. The people and places where the activities are carried out are connected by technical information networks, whether for the acquisition of raw materials, marketing or distribution.

During the Covid-19 pandemic, Joana said she worked making masks. In this case, online sales were decisive for the continuity of the business and the maintenance of the people who work with her, especially the seamstress.

As an example of the formation of an urban economy of migration tied to the upper circuit of the economy, there are investments of African migrants bound to the company Angola Cables S.A. That is a multinational company founded in 2009 with headquarters in the city of Luanda/Angola and submarine cable hubs in the cities of Miami/United States, Lisbon/Portugal, Praia/Cape Verde and Fortaleza/Brazil.

The company arrived in Fortaleza in 2016. An office was set up in Praia do Futuro, staffed by skilled professionals from Angola; hence, it is a migration in which the people involved are highly skilled and dedicated in implementing an innovation project. The project in question is the installation of the SACS (South Atlantic Cable System) - the first submarine fiber optic cable to be installed in South America, whose lenght is 6,000 km and connects the cities of Luanda and Fortaleza. The investment is made entirely by the company. The second investment in Brazil is through the installation of the Monet Cable that connects the cities of Santos (state of São Paulo), Fortaleza (state of Ceará) and Miami/USA. Such investment is being made through a consortium involving the companies Angola Cables (Angola), Google (United States), Antel (Uruguay) and Algar Telecom (Brazil).

The empirical examples presented compose production and commercialization circuits that are formed as a result of migration and which contribute to the emergence of an urban migration economy. They are tied to the lower circuit of the economy (as Dio - a street vendor); to the upper marginal circuit of the economy (as Joana's activities) and the upper circuit of the urban economy (as the Angola Cables S.A. company). The urban economy of migration which manifests itself in the lower and upper marginal circuit is not presented as an example of wealth accumulation, but it does guarantee the maintenance of people - both in the territory of migration and in the territory of origin.

These are examples of a globalization structured by those below (Portes, 1999; Tarrius, 2002) or globalization of the poor (Choplin, Pliez, 2015) that coexists with the globalization of transnational corporations (Santos, 2003; Doron, 2022).

Furthermore, as they are products that exalt elements of culture, they contribute to the rescue of identities, creation of belonging and integration between Africans and Brazilians who consume what is produced, making the urban economy of migration bring elements of an ethnic economy ( Ma Mung, 1996 ; Missaoui , Tarrius, 2013).

**Final Remarks**

The migratory flows intertwining Brazil since the beginning of the 21st century are both internal and international. Of the international flows: students, refugees, migrants with low or no professional qualifications and skilled migrants (investors) prevail. Investors include: industry; agribusiness; land purchase; development; construction industry; real estate; hotels; imports and exports. Investments of micro, small and medium-sized scales: commerce and services.

Investments are mainly made in large cities. In the Northeast, the cities of Salvador, Recife and Fortaleza stand out as the cities with the highest volume of international migrants with an investment profile. In Ceará, investments are primarily made in Fortaleza, but the municipalities on the coast and in the "countryside" of the state are also highlighted, playing a central role in Ceará's urban network.

Investors, with the permissiveness of internal agents, "normalize" and "use" the territory, producing their own territorialities and shaping the urban economy of migration. In the cities of Ceará, there are multiple territorialities that materialize in the landscape according to each economic and cultural profile of the migrants, especially international ones.

Our study of international migrations not only identifies the places of departure and arrival, but also seeks to understand the spatial, temporal and social relationships that involve migrants in the spaces in which they cross, settle, move, circulate, transit, stay - transform and create their territorialities.

Throughout their migration trajectory, differently from previous periods, migrants are involved in activities linked to the formal labor market, but a considerable number are self-employed in order to maintain their social reproduction needs in the migration territory. These activities allow the construction of cooperation, territorialities and an urban economy of migration that can be inserted into the upper, upper marginal or lower circuit of the urban economy.

In this text, our priority is to present aspects of the urban economy of migration of Africans who work in the cities of Fortaleza and its Metropolitan Region. The activities implemented by the migrants investigated have the potential to energize a local scale - the manifestation of work and life and the scale of relations - with the territory of origin and with several subjects involved in the circuits that constitute the activity accomplished.

In summary, given the work carried out, we suggest that a broad academic discussion is necessary that strives to explain the configuration and content of migration in the 21st century based on the activities carried out by migrants. In addition to theoretical reflection, we intend that the discussions held can support the construction of public policies with a view to promoting the structuring of an urban economy of popular migration, inclusive, dynamic and autonomous.

*Denise Cristina Bomtempo* - ORCID: http://orcid.org/0000-0002-0720-2110

**Acknowledgment:**

To the National Council for Scientific and Technological Development (CNPq) Brazil. Notice: CNPq/MCTI/FNDCT No. 18/2021 - Track A - Emerging Groups. Process: 422880/2021-3, for financing part of the activities carried out in the project entitled: FORCED MIGRATION IN THE NORTHEAST REGION OF BRAZIL: NETWORKS, CIRCULARITIES AND TERRITORIALITIES, which culminated, among others, in the results presented in this text.